\documentclass[fleqn,11pt]{article}

\usepackage{amsmath,amsfonts,amssymb}
\usepackage{color}
\usepackage{geometry}
\newtheorem{theorem}{Theorem}
\newtheorem{lemma}[theorem]{Lemma}

\newtheorem{proposition}{Proposition}

\newcommand{\R}{\mathbb{R}}

\newcommand{\N}{\mathbb{N}}

\newcommand{\cqfd}
{%
\mbox{}%
\nolinebreak%
\hfill%
\rule{2mm}{2mm}%
\newline
\newline
}

\title{Well-posedness of the Cauchy problem for a space-dependent anyon Boltzmann equation.}
\author{Leif ARKERYD and Anne NOURI\\
\\Mathematical Sciences, 41296 G\"oteborg, Sweden,\\
arkeryd@chalmers.se\\
Aix-Marseille University, CNRS, Centrale Marseille, I2M UMR 7373, 13453 Marseille, France,\\
anne.nouri@univ-amu.fr}

\date{}

\begin{document}

\maketitle

{\noindent \bf Abstract.}\hspace{0.1in}
 A fully non-linear kinetic Boltzmann equation for anyons is studied in a periodic 1d setting with large initial data. Strong $L^1$ solutions are obtained for the Cauchy problem. The main results  concern global existence, uniqueness and stabililty.

\footnotetext[2]{Key words; anyon, Haldane statistics, low temperature kinetic theory, quantum Boltzmann equation.}
\section{Anyons and the Boltzmann equation.}
Let us first recall the definition of anyon. Consider the wave function $\psi(R,\theta,r,\varphi)$ for two identical particles with center of mass coordinates $(R,\theta)$ and relative coordinates $(r,\varphi)$. Exchanging them,
$\varphi\rightarrow \varphi+\pi$, gives a phase factor $e^{2\pi i}$  for bosons and $e^{\pi i}$ for fermions. In three or more dimensions those are all possibilities. Leinaas and Myrheim proved in 1977 \cite{LM}, that in one and two dimensions any phase factor is
possible in the particle exchange. This became an important topic after the first experimental confirmations in the early 1980-ies, and Frank Wilczek in analogy with the terms bos(e)-ons and fermi-ons coined the name any-ons  for the new quasi-particles with any phase. Anyon quasi-particles with e.g. fractional electric charge, have since been observed in various types of experiments.\\
By moving to a definition in terms of a generalized Pauli exclusion principle, Haldane \cite{H} extended this  to a fractional exclusion statistics valid for any dimension, and coinciding with the anyon definition in the one and two dimensional cases. Haldane statistics has also been realized for neutral fermionic atoms at ultra-low temperatures in three dimensions \cite{BMB}. Wu later derived \cite{W} occupation-number distributions for ideal gases under Haldane statistics by counting states under the new fractional exclusion principle.
From the number of quantum states  of $N$ identical particles occupying $G$ states being
\begin{eqnarray*}
\frac{(G+N-1)!}{ N!(G-1)!} \hspace{.5cm} {\rm and}\hspace{.5cm} \frac{G!}{N!(G-N)!}
\end{eqnarray*}
in the boson resp. fermion cases, he derived the interpolated number of quantum states for the fractional exclusions to be
\begin{eqnarray}
\frac{(G+(N-1)(1-\alpha))!}{N!(G-\alpha N-(1-\alpha))!}\hspace{.1cm},\hspace{1cm} 0<\alpha<1.
\end{eqnarray}
He then obtained for ideal gases the equilibrium statistical distribution
\begin{eqnarray}
\frac{1}{w(e^{(\epsilon-\mu)/T})+\alpha}\hspace{.1cm},
\end{eqnarray}
where $\epsilon$ denotes particle energy, $\mu$ chemical potential, $T$ temperature, and the function $w(\zeta)$ satisfies
\begin{eqnarray*}
w(\zeta)^\alpha(1+w(\zeta))^{1-\alpha}=\zeta\equiv e^{(\epsilon-\mu)/T}.
\end{eqnarray*}
In particular $w(\zeta)=\zeta-1$ for $\alpha=0$ (bosons) and $w(\zeta)=\zeta$ for $\alpha=1$ (fermions).\\
\hspace{1cm}\\
In elastic pair collisions, the velocities $(v,v_*)$ before and $(v',v'_*)$ after a collision are related by
\begin{eqnarray*}
v'= v-n[(v-v_*)\cdot n],\quad v'_*=v_*+n[(v-v_*)\cdot n],\quad n\in S^{d-1}.
\end{eqnarray*}
This preserves mass, linear momentum, and energy in Boltzmann type collision operators.
We shall write $ f= f(v),\quad f_*= f(v_*),\quad f'= f(v'),\quad f'_*= f(v'_*)$.
An important question for gases with fractional exclusion statistics, is how to calculate their transport properties, in particular how the Boltzmann equation
 \begin{equation*}
\partial _tf+v\cdot\bigtriangledown_xf= Q(f)
\end{equation*}
 gets modified.
An answer was given by Bhaduri, Bhalerao, and Murthy \cite{BBM} by generalizing to anyons the filling factors $F(f)$ from the fermion and boson cases, $F(f)=(1+\eta f)$, $\eta=\mp 1$, {and by inductive reasoning obtaining as anyon filling factors $F(f)= (1-\alpha f)^{\alpha}(1+(1-\alpha)f)^{1-\alpha}$, $0<\alpha<1$.}\\
%The filling factors appear because quantum transitions also depend on the occupancy of the final state.
Namely, with a filling factor $F(f)$  in the collision operator $Q$, the entropy production term becomes
\begin{eqnarray*}
\int Q(f)\log\frac{f}{F(f)}dv\hspace{.1cm},
\end{eqnarray*}
which for equilibrium implies
\begin{eqnarray*}
\frac{f'}{F(f')}\frac{f'_*}{F(f'_*)}=\frac{f}{F(f)}\frac{f_*}{F(f_*)}\hspace{.1cm}.
\end{eqnarray*}
{Using conservation laws and properties of the Cauchy equation, one concludes that in equilibrium $\frac{f}{F(f)}$ is a Maxwellian.  Inserting Wu's equilibrium (1.2) for $f$ and taking the quotient Maxwellian as $e^{-(\epsilon-\mu)/T}$, this gives}
\begin{eqnarray*}
f=\frac{1}{w(e^{(\epsilon-\mu)/T})+\alpha}\hspace{.1cm},\quad \quad F(f)=f e^{(\epsilon-\mu)/T}=\frac{e^{(\epsilon-\mu)/T}}{w(e^{(\epsilon-\mu)/T})+\alpha}\hspace{.1cm}.
\end{eqnarray*}
In particular in the fermion and boson cases,
\begin{eqnarray*}
f=\frac{1}{e^{(\epsilon-\mu)/T}-\eta},\quad \quad F(f)=\frac{e^{(\epsilon-\mu)/T}}{e^{(\epsilon-\mu)/T}-\eta},
\hspace{.1cm}\eta=\mp1.
\end{eqnarray*}
This is consistent with taking an interpolation between the fermion and boson factors as general filling factor, $F(f)= (1-\alpha f)^{\alpha}(1+(1-\alpha)f)^{1-\alpha}$, $0<\alpha<1$.
{It gives the collision operator $Q$ of \cite{BBM} for Haldane statistics,}
\begin{eqnarray}
Q(f)(v)=Q^+(f)-Q^-(f)= \int_{I\! \!R^d \times S^{d-1}}B(|v-v_*|,\omega)
 [f'f'_*F(f)F(f_*)-ff_*F(f')F(f'_*)] dv_*d\omega.\hspace{.1cm}
\end{eqnarray}
Here $d\omega$ corresponds to the Lebesgue probability measure on the $(d-1)$-sphere. The collision kernel $B(z,\omega)$ in the variables $(z,\omega)\in I\! \!R^d \times \mathbb{S}^{d-1}$ is positive, locally integrable, and only depends on $|z|$ and $|(z,\omega)|$.
See [2] for a further discussion of the kernel $B$.\\
The anyon Boltzmann equation for $0<\alpha<1$ retains important properties from the Fermi-Dirac case, but it has so far not been validated from basic quantum theory.
In the filling factor $F(f)= (1-\alpha f)^{\alpha}(1+(1-\alpha)f)^{1-\alpha}$, $0<\alpha<1$, the factor $(1-\alpha f)^\alpha$ requires the value of $f$ to be between $0$ and $\frac{1}{\alpha}$. This is formally preserved by the equation, since the gain term vanishes for $f=\frac{1}{\alpha}$, making the $Q$-term (1.3) and the derivative left hand side of the Boltzmann equation negative there. And the derivative equals the positive gain term for $f=0$, where the loss term vanishes.
$F$ is concave with maximum value one at $f=0$ for $\alpha \geq \frac{1}{2}$, and maximum value
$(\frac{1}{\alpha}-1)^{1-2\alpha}>1$ at $f= \frac{1-2\alpha}{\alpha(1-\alpha)}$ for $\alpha <\frac{1}{2}$. The collision  operator vanishes identically for the equilibrium distribution functions obtained by Wu, but for no other functions.\\
The Boltzmann equation for the limiting cases, representing boson statistics ($\alpha=0$)
and fermion statistics ($\alpha=1$), was introduced by Nordheim \cite{N}  in 1928. Here the quartic terms in the collision integral cancel, which is used in the analysis. General existence results for the space-homogeneous isotropic boson large data case were obtained in \cite{Lu1}, followed by a number of other papers, e.g. \cite{EMV}, \cite{Lu2}, \cite{Lu3}, \cite{Lu4}, and for the space-dependent case near equilibrium in \cite{R}.  In the space-dependent fermion case general existence results were obtained in \cite{D} and \cite {L}.\\
For $0<\alpha<1$ there are { no cancellations} in the collision term. Moreover, the Lipschitz continuity of the collision term is replaced by a {weaker H\"older continuity} near $f=\frac{1}{\alpha}$. The space-homogeneous initial value problem for the Boltzmann equation with Haldane statistics is
\begin{eqnarray}
\frac{ df}{dt}=Q(f),\quad f(0,v)=f_0(v).
\end{eqnarray}
Because of the filling factor $F$, the range for the initial value $f_0$ should belong to  $[0,\frac{1}{\alpha}]$, which is also formally preserved by the equation. A good control of $\int f(t,x,v)dv$, which in the space-homogeneous case is given by the mass conservation, can be used to keep $f$ uniformly away from $\frac{1}{\alpha}$, and $F(f)$ Lipschitz continuous. That was a basic observation behind the existence result for the space-homogeneous anyon Boltzmann equation.
\setcounter{theorem}{0}
\begin{proposition}
{\rm [1]} Consider the space-homogeneous equation (1.4) with velocities in $I\! \!R^{d}$, $d\geq 2$ and for hard force kernels with
\begin{eqnarray}
0<B(z,\theta)\leq C|z|^{\beta} |\sin \theta \cos \theta |^{d-1},
\end{eqnarray}
where {$\frac{-\pi}{2}\leq\theta\leq\frac{\pi}{2}$}, $0<\beta \leq 1$, $d>2$, and $0<\beta<1$, $d=2$. Let the initial value $f_0$ have finite mass and energy. If $0<f_0\leq\frac{1}{\alpha}$ and ${\rm ess\hspace{.1cm} sup}(1+|v|^s)f_0 <\infty$ for $s=d-1+\beta$, then the initial value problem for (1.4)
has a strong solution in the space of functions continuous from $t\geq 0$ into $L^1\cap L^{\infty}$, which conserves mass and energy, and for $t_0>0$ given, has ${\rm ess\hspace{.1cm}sup}_{v,t\leq t_0}|v|^{s'}f(t,v)$ bounded,where $s'=\min(s,\frac{2\beta(d+1)+2}{d})$.
\end{proposition}
In this proposition, stronger limitations on $B$ would allow for weaker conditions on the initial value $f_0$.
The proof implies stability; given a sequence of positive initial values  $(f_{0n})_{n\in\mathbb{N}}$ with
\begin{eqnarray*}
\sup_n {\rm ess\hspace{.1cm}sup}\hspace{.2cm} f_{0n}(v) <\frac{1}{\alpha},
\end{eqnarray*}
and converging in $L^1$  to $f_0$, there is a subsequence of the solutions converging in $L^1$ to a solution with initial value $f_0$.\\
\newpage
\setcounter{equation}{0}
\setcounter{theorem}{0}
\setcounter{equation}{0}
\setcounter{theorem}{0}
\section{The main results.}
The present paper considers the space-dependent {anyon Boltzmann equation in a slab}.
Anyons only exist in one and two dimensions. {The proof in this paper uses an estimate for the Bony functional in one space dimension, which due to the filling factor $F(f)$, is restricted to the anyon case $v\in\R^2$.}
%and their velocities are restricted  to $v\in\mathbb{R}^2$.
{For $cos\theta = n\cdot\frac{v-v_*}{|v-v_*|}$, the kernel $B(|v-v_*|, \theta )$ is assumed measurable with
\begin{equation}\label{hyp1-B}
0\leq B\leq B_0,
\end{equation}
for some $B_0>0$. It is also assumed for some $\gamma, \gamma',{c_B}>0$, that
\begin{equation}\label{hyp2-B}
B(|v-v_*|, \theta )=0 \hspace*{0.05in}\text{for}\hspace*{0.05in}   |cos\theta |<\gamma',\quad
\text{for}\hspace*{0.05in}  1-|cos\theta|<\gamma',\quad  \text{and for   } |v-v_*|< \gamma,
\end{equation}
 and that
 \begin{equation}\label{hyp3-B}
\int B(|v-v_*|, \theta )d\theta \geq c_B>0\quad  \text{for   }|v-v_*|\geq \gamma .
\end{equation}
}The initial datum $f_0(x,v)$, periodic in $x$, is assumed to be a measurable function with values in $]0,\frac{1}{\alpha }] $,
and such that
\begin{eqnarray}\label{Hyp-f0}
(1+|v|^2)f_0(x,v) \in L^1([ 0,1] \times \R ^2), \hspace{.1cm}
\int \sup_{x\in[0,1]} f_0(x,v)dv=c_0 <\infty,\hspace{.1cm}
\inf_{x\in[0,1]}f_0(x,v)>0,\hspace*{0.1cm}\text{a.a.}v\in \R ^2 .\hspace{.1cm}
\end{eqnarray}
With $v_1$ denoting the component of $v$ in the $x$-direction, consider for functions periodic in $x$, the initial value problem
\begin{equation}\label{f}
\partial _tf(t,x,v)+v_1\partial _xf(t,x,v)= Q(f)(t,x,v),\quad f(0,x ,v )= f_0(x,v),\quad{(t,x,v)\in \R _+\times [ 0,1] \times \R ^2.}
\end{equation}
The main result of the present paper is the following theorem.
\setcounter{theorem}{0}
\begin{theorem}
\hspace*{0.1in}\\
There exists a strong solution $f\in\mathcal{C}([0,\infty [;L^1([0,1]\times\R^2))$ of (\ref{f}) with $0<f(t,.)<\frac{1}{\alpha}$ for $t>0$. There is $t_m>0$ such that for any $T>t_m$, there is $\eta_T>0$ so that $f\leq \frac{1}{\alpha}-\eta_T$  for $t_m\leq t\leq T$.\\
 The solution is unique and stable in the $L^1$-norm on each interval {of time $[ 0,T] $}. \\
 It conserves mass, first $v$-moments and energy.
\end{theorem}
\underline{\bf Remarks.}\\
The above results seem to be new also in the fermion case where $\alpha=1$.\\
The approach in the paper can also be used to obtain regularity results. \\
The control of $\int f(t,x,v)dv$ is in the present space-dependent setting is non-trivial.\\
The asymptotic behaviour of the solution, not considered in this paper, is related to an entropy for (\ref{f}),
\begin{eqnarray*}
\int \Big(f\log f +(\frac{1}{\alpha}-f)\log (1-\alpha f)^\alpha -(\frac{1}{1-\alpha}+f)\log (1+(1-\alpha)f)^{1-\alpha}\Big)dxdp.
\end{eqnarray*}
\\
 \\
An open problem is the behaviour of (\ref{f}) beyond the anyon frame, i.e. for higher $v$-dimensions under Haldane statistics. It seems likely that a close to equilibrium approach as in the classical case, could work with fairly general kernels $B$ for close to equilibrium initial values $f_0$ with some regularity and strong decay conditions for large velocities.
Any progress on the large data case in several space-dimensions under Haldane statistics would be quite interesting.\\
\\
The lack of Lipschitz continuity of $F(f)$ when $f$ is in a neighborhood of $\frac{1}{\alpha }$ requires some care. Since the gain term vanishes when $f=\frac{1}{\alpha}$ and the derivative becomes negative there, $f$ should start decreasing before reaching this value. The proof that this takes place uniformly over phase-space and approximations, is based on a good control of $\int f(t,x,v)dv$ in the integration of the gain and loss parts of $Q$. That is a main topic in Section 3 together with the study of a family of approximating equations with large velocity cut-off. Based on those results and using the Lipschitz continuity of $F(.)$ away from $\frac{1}{\alpha}$, in Section 4 contraction mapping techniques prove the well-posedness of the problem, when the initial value $f_0$ stays uniformly away from $\frac{1}{\alpha}$. That restriction  is removed by a local initial value analysis, which only assumes H\"older continuity of $F(.)$.
\setcounter{equation}{0}
\setcounter{theorem}{0}
\section{Approximations and control of mass density.}
\setcounter{equation}{0}
\setcounter{theorem}{0}
For any $j\in \N ^*$, denote by $\psi_j$, the cut-off function with
\[ \begin{aligned}
&\psi_j(r)=0\quad \text{if   }r>j
&\text{and}\quad \psi_j(r)= 1\quad \text{if  }
r\leq j,
\end{aligned}\]
and set
\[ \begin{aligned}\chi_j(v,v_*,v',v'_*)=\psi_j(|v| ) \psi_j(|v_*| ) \psi_j(|v'| ) \psi_j(|v'_*| ).
\end{aligned}\]
Let $F_j$ be defined on $[ 0,\frac{1}{\alpha }] $ by
\begin{eqnarray*}
 F_j(y)= \frac{1-\alpha y}{(\frac{1}{j} +1-\alpha y)^{1-\alpha }} (1+(1-\alpha )y)^{1-\alpha }.
\end{eqnarray*}
Denote by $Q_j$ {(resp. $Q_j^+$), the operator
\[ \begin{aligned}
Q_j(f)(v):= \frac{1}{\pi }\int B(|v-v_*|,{\theta} )\chi _j(v,v_*,v',v'_*)\Big( f^\prime f^\prime _*F_j(f)F_j(f_*)-ff_*F_j(f^\prime )F_j(f^\prime _*)\Big) dv_*d{\theta } ,\\
{(\text{resp. its gain part    }Q_j^+(f)(v):= \frac{1}{\pi }\int B(|v-v_*|,{\theta } )\chi _j(v,v_*,v',v'_*)f^\prime f^\prime _*F_j(f)F_j(f_*)dv_*d{\theta } ).}
\end{aligned}\]
For $j \in \N ^*$, let a mollifier $\varphi _j$ be defined by  $\varphi _j(x,v)= j^3\varphi (jx,jv)$, where
\begin{eqnarray*}
\varphi \in C_0^\infty (\R ^3),\quad support (\varphi )\subset [ 0,1] \times \{ v\in \R ^2;\lvert v\rvert \leq 1\},\quad \varphi \geq 0,\quad \int \varphi (x,v)dxdv= 1.
\end{eqnarray*}
Let $f_{0,j}$ be the restriction to $[ 0,1] \times \{v; \lvert v\rvert \leq j\}$ of $\big( \min \{f_0, \frac{1}{\alpha }-\frac{1}{j} \} \big) \ast \varphi _j$. \\
The following lemma concerns a corresponding approximation of (2.5).\\
%\textcolor{red}{Since $\eta _j$ depends on $T$, I have reformulated the statement of Lemma 3.1.}
%
%
% Lemma 3.1
%
\begin{lemma}
{For $T>0$, there is a unique solution $f_j\in {C}([ 0,T] \times [ 0,1] ;L^1( \{v; \lvert v\rvert \leq j\} ))$ to
\begin{equation}\label{eq-f_j}
\partial _tf_j+v_1\partial _xf_j= Q_j(f_j),\quad f_j(0,\cdot ,\cdot )= f_{0,j},
\end{equation}
with values in $ ] 0,\frac{1}{\alpha }-\eta _j] $, for some $\eta _j>0$.} It conserves mass, first moment and energy.
\end{lemma}
\underline{Proof of Lemma 3.1.}\\
Let $T>0$ be given. We shall first prove by contraction that for $T_1>0$ and small enough, there is a unique solution
\begin{eqnarray*}
f_{\epsilon ,j}\in C([ 0,T_1] \times [ 0,1] ; L^1( \{ v; \lvert v\rvert \leq j\} ))\cap \{f; f\in [ 0,\frac{1}{\alpha } ] \}
\end{eqnarray*}
to (\ref{eq-f_j}). Let the map $\mathcal{C}$ be defined on {periodic in $x$ functions in} $C ([ 0,T] \times [ 0,1] ;L^1( \{ v; \lvert v\rvert \leq j\} )) \cap \{f; f\in [ 0,\frac{1}{\alpha } ] \}$ by $\mathcal{C}(f)= g$, where
\[ \begin{aligned}
&\partial _tg +v_1\partial _xg = \frac{1}{\pi }(1-\alpha g)\Big( \frac{1+(1-\alpha )f}{\frac{1}{j}+1-\alpha f}\Big) ^{1-\alpha }\int B\chi _j{f}^\prime {f}^\prime _*F_j({f}_*)dv_*d{\theta } -\frac{g}{\pi }\int B\chi _j{f}_*F_j(f^\prime )F_j(f^\prime _*)dv_*d{\theta }  ,\\
&g(0,\cdot ,\cdot )= f_{0,j}.
\end{aligned}\]
It follows from the linearity of the previous partial differential equation that it has a unique {periodic} solution $g$ in $C ([ 0,T] \times [ 0,1] ;L^1( \{ v; \lvert v\rvert \leq j\} )) $. For $f$ with values in $ [ 0,\frac{1}{\alpha } ] $, $g$ takes its values in $ ] 0,\frac{1}{\alpha } ] $. Indeed, denoting by $g^\sharp (t,x,v)= g(t,x+tv_1,v)$,
\begin{eqnarray*}
g^\sharp (t,x,v)\geq &f_{0,j}(x+tv_1,v)e^{-\int _0^t\bar{\sigma }_f^\sharp (r,x,v)dr}>0,
\end{eqnarray*}
and
\[ \begin{aligned}
(1-\alpha g)^\sharp (t,x,v)&= (1-\alpha f_{0,j})(x+tv_1,v)e^{-\int _0^t\tilde{\sigma }_f^\sharp (r,x,v)dr}\\
&+\frac{\alpha }{\pi }\int _0^t\Big( g\int B\chi _jf_*F_j(f^\prime )F_j(f^\prime _*)dv_*{d\theta} \Big) ^\sharp (s,x,v)e^{-\int _s^t\tilde{\sigma }_f^\sharp (r,x,v)dr}ds\\
&\geq (1-\alpha f_{0,j})(x+tv_1,v)e^{-\int _0^t\tilde{\sigma }_f^\sharp (r,x,v)dr}\geq 0.
\end{aligned}\]
Here,
\[ \begin{aligned}
&\bar{\sigma }_f:= \frac{\alpha }{\pi }\Big( \frac{1+(1-\alpha )f)}{\frac{1}{j}+1-\alpha f}\Big) ^{1-\alpha }\int B\chi _jf^\prime f^\prime _*F_j(f_*)dv_*d{\theta } +\frac{1}{\pi }\int B\chi _jf_*\tilde{F}_{\epsilon ,j}(f^\prime )\tilde{F}_{\epsilon ,j}(f^\prime _*)dv_*d{\theta }  ,\\
&\tilde{\sigma }_f:= \frac{\alpha }{\pi }\Big( \frac{1+(1-\alpha )f)}{\frac{1}{j}+1-\alpha f}\Big) ^{1-\alpha }\int B\chi _jf^\prime f^\prime _*F_j(f_*)dv_*d{\theta } .
\end{aligned}\]
$\mathcal{C}$ is a contraction on $C([0,T_1] \times [ 0,1] ; L^1(\{ v;\lvert v\rvert \leq j\} ))\cap \{f; f\in [ 0,\frac{1}{\alpha } ] \}$, for $T_1>0$ small enough only depending on $j$, since the derivative of the map $F_j$ is bounded on $[ 0,\frac{1}{\alpha }] $. Let $f_j$ be its fixed point, i.e. the solution of (\ref{eq-f_j}) on $[ 0,T_1] $. The argument can be repeated and the solution can be continued up to $t=T$. By the exponential form for $f_j$ (resp. $1-\alpha f_j$)
\[ \begin{aligned}
f_j^\sharp (t,x,v)&\geq f_{0,j}(x,v)e^{-\int _0^t\bar{\sigma }_{f_j}^\sharp (r,x,v)dr}>0,
\quad t\in [ 0,T]
,\hspace{0.03in}x\in [ 0,1] ,\hspace{0.03in}\lvert v\rvert \leq j,
\end{aligned}\]
(resp.
\[ \begin{aligned}
(1-\alpha f_j)^\sharp (t,x,v)&\geq (1-\alpha f_{0,j})(x+tv_1,v)e^{-\int _0^t\tilde{\sigma }_{f_j}^\sharp (r,x,v)dr}\\
&\geq \frac{1}{je^{cj^3T}},
\quad t\in [ 0,T]
,\hspace{0.03in}x\in [ 0,1] ,\hspace{0.03in}\lvert v\rvert \leq j).
\end{aligned}\]
Consequently, for some $\eta_j>0$, there  is a {periodic in $x$} solution $f_j\in C([ 0,T] \times [ 0,1] ;L^1( \{ v;\lvert v\rvert \leq j\} ))$ to (\ref{eq-f_j})  with values in $] 0, \frac{1}{\alpha }-{\eta_j}]$. \\
If there were another nonnegative local solution $\tilde{f}_{j}$ to (\ref{eq-f_j}), defined on $[ 0,T^\prime ] $ for some $T^\prime \in ] 0,T] $, then by the exponential form it would stay below $\frac{1}{\alpha}$. The difference $f_j-\tilde{f}_j$ would for some constant $c_{T^\prime}$ satisfy
\begin{eqnarray*}
\int \lvert (f_j -\tilde{f}_j)^\sharp (t,x,v)\rvert dxdv\leq c_{T^\prime }\int _0^t\lvert (f_j-\tilde{f}_j)^\sharp (s,x,v)\rvert dsdxdv,\hspace*{0.03in}t\in [ 0,T^\prime ],\quad  (f_j-\tilde{f}_j)^\sharp (0,x,v)= 0,
\end{eqnarray*}
implying that the difference {would be} identically zero on $[ 0,T^\prime ] $. Thus $f_j$ is the unique solution on $[ 0,T] $ to (\ref{eq-f_j}), and has its range contained in $] 0,\frac{1}{\alpha }-\eta _j] $.\\
Moreover, $f_j\in W^{1,1}([ 0,T] \times [ 0,1] ;L^1( \{ v;\lvert v\rvert \leq j\} ))$. Indeed, $\partial _xf_j$ satisfies
\begin{eqnarray}
\partial _t(\partial _xf_j)+v_1\partial _x(\partial _xf_j)+\sigma _j\partial _xf_j= \frac{1}{\pi }(1-\alpha f_j)\partial _x\Big( (\frac{1+(1-\alpha f_j)}{\frac{1}{j}+1-\alpha f_j})^{1-\alpha }\int B\chi _jf_j^\prime f^\prime _{j*}F_j(f_{j*})dv_*d\theta \Big) \nonumber \\
-\frac{f_j}{\pi }\partial _x\int B\chi _j f_{j*}F_j(f_j^\prime )F_j(f^\prime _{j*})dv_*d\theta ,\quad \\
\partial _xf_j(0,\cdot ,\cdot )= \partial _xf_{0,j},\hspace*{4.75in}
\end{eqnarray}
where
\[ \begin{aligned}
&\sigma _j:= \frac{\alpha }{\pi }(\frac{1+(1-\alpha f_j)}{\frac{1}{j}+1-\alpha f_j})^{1-\alpha }\int B\chi _jf_j^\prime f^\prime _{j*}F_j(f_{j*})dv_*d\theta +\frac{1}{\pi }\int B\chi _jf_{j*}F_j(f_j^\prime )F_j(f^\prime _{j*})dv_*d\theta .
\end{aligned}\]
Using the exponential form of $\partial _xf_j$, multiplying it by $sgn(\partial _xf_j)$, integrating the resulting equation on $[ 0,1] \times \{ v;\lvert v\rvert \leq j\} $ and using a Gronwall argument leads to a $j$-dependent bound for $\int \lvert \partial _xf_j(t,x,v)\rvert dxdv$ on $[ 0,T] $. Hence, also from (\ref{eq-f_j}) and the bounded domain of integration of $v$, $\partial _tf_j$ also belongs to $L^\infty (0,T; L^1([ 0,1] \times \{ v;\lvert v\rvert \leq j\} ))$.  \cqfd
\hspace*{0.1in}\\
The remaining part of this section is devoted to obtaining a uniform control with respect to $j\in \N ^*$ of
\begin{eqnarray*}
\int \sup _{t>0,\hspace*{0.02in}x\in [ 0,1] }f_j^\sharp (t,x,v)dv.
\end{eqnarray*}
It relies on the following four lemmas,  where the first is an estimate of the Bony functionals,
\begin{eqnarray*}
\bar{B}_j(t):= \int_0^1\int |v-v_*|^2 B{\chi}_jf_jf_{j*}F_j(f'_j)F_j(f'_{j*})dvdv_*d\theta dx,\quad t\geq 0.
\end{eqnarray*}
\begin{lemma}
\hspace*{0.1in}\\
{For $T>0$ it holds that
\begin{eqnarray*}
\int_0^{T }\bar{B}_j(t)dt\leq c'_0(1+T),\quad j\in \N ^*,
\end{eqnarray*}
with $c'_0$ only depending on $\int f_0(x,v)dxdv$ and on $\int |v|^2f_0(x,v)dxdv$.}
\end{lemma}
\underline{Proof of Lemma 3.2.} \\
%\textcolor{red}{New proof, from the Cercignani-Illner paper}\\
Denote $f_j$ by $f$ for simplicity. The proof {is} an extension of the classical one (cf \cite{B}, \cite{CI}), as follows. The integral over time of the momentum $\int v_1f(t,0,v)dv$ (resp. the momentum flux \\
$\int v_1^2f(t,0,v)dv$ ) is first controlled. Let $\beta \in C^1([ 0,1] )$ be such that $\beta (0)= -1$ and $\beta (1)= 1$. Multiply (\ref{eq-f_j}) by $\beta (x)$ (resp. $v_1\beta (x)$ ) and integrate over $[ 0,t] \times [ 0,1] \times \R ^2$. It gives
\[ \begin{aligned}
\int _0^t\int v_1f(\tau ,0,v)dvd\tau = \frac{1}{2}\big( \int \beta (x)f_0(x,v)dxdv&-\int \beta (x)f(t,x,v)dxdv\\
&+\int _0^t\int \beta ^\prime (x)v_1f(\tau ,x,v)dxdvd\tau\big) ,
\end{aligned}\]
\Big( resp.
\[ \begin{aligned}
\int _0^t\int v_1^2f(\tau ,0,v)dvd\tau = \frac{1}{2}\big( \int \beta (x)v_1f_0(x,v)dxdv&-\int \beta (x)v_1f(t,x,v)dxdv\\
&+\int _0^t\int \beta ^\prime (x)v_1^2f(\tau ,x,v)dxdvd\tau\big) \Big) .
\end{aligned}\]
Consequently, using the conservation of mass and energy of $f$,
\begin{align}\label{bony-1}
\lvert \int _0^t\int v_1f(\tau ,0,v)dvd\tau \rvert +\int _0^t\int v_1^2f(\tau ,0,v)dvd\tau \leq c(1+t).
\end{align}
Let
\begin{eqnarray*}
\mathcal{I}(t)= \int _{x<y}(v_1-v_{*1})f(t,x,v)f(t,y,v_*)dxdydvdv_*.
\end{eqnarray*}
It results from
\begin{eqnarray*}
\mathcal{I}'(t)= -\int (v_1-v_{*1})^2f(t,x,v)f(t,x,v_*)dxdvdv_*+2\int v_{*1}(v_{*1}-v_1)f(t,0,v_*)f(t,x,v)dxdvdv_*,
\end{eqnarray*}
and the conservations of the mass, momentum and energy of $f$ that
\begin{align}\label{bony-2}
&\int _0^t \int_0^1 \int (v_1-v_{*1})^2 f(s,x,v)f_*(s,x,v_*)dvdv_*dxds\nonumber \\
&\leq 2\int f_0(x,v)dxdv\int \lvert v_1\rvert f_0(x,v)dv+ 2\int f(t,x,v)dxdv\int \lvert v_1\rvert f(t,x,v)dxdv\nonumber \\
&+2\int _0^t\int v_{*1}(v_{*1}-v_1)f(\tau ,0,v_*)f(\tau ,x,v)dxdvdv_*d\tau \nonumber \\
&\leq 2\int f_0(x,v)dxdv\int (1+\lvert v\rvert ^2)f_0(x,v)dv+ 2\int f(t,x,v)dxdv\int (1+\lvert v\rvert ^2) f(t,x,v)dxdv\nonumber \\
&+2\int _0^t(\int v_{*1}^2f(\tau ,0,v_*)dv_*)d\tau\int f_0(x,v)dxdv-2\int _0^t(\int v_{*1}f(\tau ,0,v_*)dv_*)d\tau\int  v_1f_0(x,v)dxdv\nonumber \\
&\leq c\Big( 1+\int _0^t\int v_1^2f(\tau ,0,v)dvd\tau +\lvert \int _0^t\int v_1f(\tau ,0,v)dv\rvert \Big) . \nonumber
\end{align}
And so, by (\ref{bony-1}),
\begin{equation}\label{bony-3}
\int _0^t \int_0^1 \int (v_1-v_{*1})^2 f(\tau ,x,v)f(\tau ,x,v_*)dxdvdv_*d\tau \leq c(1+t).
\end{equation}
Here, $c$ is a constant depending only on $\int f_0(x,v)dxdv$ and $\int \lvert v\rvert ^2f_0(x,v)dxdv$. \\
Denote by $u_1=\frac{\int v_1fdv}{\int fdv}$. It holds
\begin{align}\label{bony-4}
\int_0^t\int_0^1 \int (v_1-u_1)^2 B{\chi}_jff_*&F_j(f')F_j(f'_*)(s,x,v,v_*,\theta )dvdv_*d\theta dxds\nonumber \\
&\leq c\int_0^t  \int_0^1 \int (v_1-u_1)^2 ff_*(s,x,v,v_*)dvdv_* dxds\nonumber \\
&= \frac{c}{2}\int _0^t \int_0^1 \int (v_1-v_{*1})^2 ff_*(s,x,v,v_*)dvdv_*dxds\nonumber \\
&\leq c(1+t).
\end{align}
Multiply equation (\ref{eq-f_j}) for $f$  by $v_1^2$, integrate and use that $\int v_1^2Q_j(f)dv= \int (v_1-u_1)^2Q_j(f)dv$ and (\ref{bony-4}). It results
\[ \begin{aligned}
&\frac{1}{\pi }\int _0^t\int (v_1-u_1)^2B{\chi}_jf^\prime f^\prime _*F_j(f)F_j(f_*)dvdv_*d\theta dxds\\
&= \int v_1^2f(t,x,v)dxdv-\int v_1^2f_0(x,v)dxdv+\frac{1}{\pi }\int _0^t\int (v_1-u_1)^2B{\chi}_jff_*F_j(f^\prime )F_j(f^\prime _*)dxdvdv_*d\theta ds\\
&<c_0(1+t),
\end{aligned}\]
where $c_0$ is a constant only depending on $\int f_0(x,v)dxdv$ and $\int \lvert v\rvert ^2f_0(x,v)dxdv$.\\
\hspace{1cm}\\
After  a collision transform the left hand side can be written
\[ \begin{aligned}
&\frac{1}{\pi }\int _0^t\int (v'_1-u_1)^2B{\chi}_jff_*F_j(f')F_j(f'_*)dvdv_*d\theta dxds\\
&= \frac{1}{\pi }\int _0^t\int (c_1-n_1[(v-v_*)\cdot n])^2B{\chi}_jff_*F_j(f')F_j(f'_*)dvdv_*d\theta dxds,
\end{aligned}\]
where $c_1=v_1-u_1$.
Expand $(c_1-n_1[(v-v_*)\cdot n])^2$, and remove the  positive term containing $c_1^2$. \\
\\
The term containing $n_1^2[(v-v_*)\cdot n]^2$ is estimated from below.
When $n$ is replaced by an orthogonal (direct) unit vector $n_\perp $, $v^\prime $ and $v^\prime _*$ are shifted and the product $ff_*F_j(f^\prime )F_j(f^\prime _*)$ is unchanged. In $\R^2$ the ratio between the sum of the integrand factors $n_1^2[(v-v_*)\cdot n]^2+ n_{\perp 1}^2[(v-v_*)\cdot n_{\perp}]^2$ and $|v-v_*|^2$, is, outside of the angular cut-off {(2.2)}, uniformly bounded from below by {$\gamma ^{\prime 2}$}. Indeed, if $\alpha $ denotes the angle between $\frac{v-v_*}{ |v-v_*|}$ and $n$,
\[ \begin{aligned}
n_1^2[\frac{v-v_*}{ |v-v_*|}\cdot n]^2+ n_{\perp 1}^2[\frac{v-v_*}{ |v-v_*|}\cdot n_{\perp}]^2&= cos^2\theta \hspace*{0.02in}cos^2\alpha +sin^2\theta \hspace*{0.02in}sin^2\alpha \\
&\geq \gamma ^{\prime 2}cos^2\alpha +\gamma ^\prime (2-\gamma^\prime )sin^2\alpha \\
&\geq\gamma ^{\prime 2},\quad \gamma ^\prime <|cos\theta | <1-\gamma ^\prime ,\quad \alpha \in [ 0,2\pi ] .
\end{aligned}\]
This is where the condition $v\in \R ^2$ is used.\\
\hspace*{0.1in}\\
That leads to the lower bound
\begin{eqnarray*}
\int_0^t\int n_1^2[(v-v_*)\cdot n]^2B{\chi}_jff_*F_j(f')F_j(f'_*)dvdv_*d\theta dxds\\
\geq {\gamma ^{\prime 2}}\int _0^t \int |v-v_*|^2B{\chi}_jff_*F_j(f')F_j(f'_*)dvdv_*d\theta dxds.
\end{eqnarray*}
And so,
\[ \begin{aligned}
&{\gamma ^{\prime 2}}\int _0^t \int |v-v_*|^2B{\chi}_jff_*F_j(f')F_j(f'_*)dvdv_*d\theta dxds\\
&\leq c_0{(1+t)}+2\int _0^t\int (v_1-u_1)n_1[(v-v_*)\cdot n]B{\chi}_jff_*F_j(f')F_j(f'_*)dvdv_*d\theta dxds\\
&\leq {c_0}{(1+t)}+2\int _0^t\int \Big( v_1(v_2-v_{*2})n_1n_2
%+p_1(p_3-p_{*3})n_1n_3
\Big) B{\chi}_jff_*F_j(f')F_j(f'_*)dvdv_*d\theta dxds,
\end{aligned}\]
since
\[ \begin{aligned}
\int &u_1(v_1-v_{*1})n_1^2B{\chi}_jff_*F_j(f')F_j(f'_*)dvdv_*d\theta dx\\
&= \int u_1(v_2-v_{*2})n_1n_2{\chi}_jBff_*F(if')F_j(f'_*)dvdv_*d\theta dx
%&= \int u_1(p_3-p_{*3})n_1n_3\bar{\chi}_jBff_*F(f')F(f'_*)dxdpdp_*d\omega \\
= \hspace*{0.01in}0,
\end{aligned}\]
by an exchange of the variables $v$ and $v_*$. Moreover, exchanging first the variables $v$ and $v_*$,
\[ \begin{aligned}
2\int _0^t&\int v_1(v_2-v_{*2})n_1n_2B{\chi}_jff_*F_j(f')F_j(f'_*)dvdv_*d\theta dxds\\
= &\int _0^t\int (v_1-v_{*1})(v_2-v_{*2})n_1n_2B{\chi}_jff_*F_j(f')F_j(f'_*)dvdv_*d\theta dxds\\
\leq &\frac{1}{\gamma ^{\prime 2}}\int _0^t\int (v_1-v_{*1})^2n_1^2B{\chi}_jff_*F_j(f')F_j(f'_*)dvdv_*d\theta dxds\\
&+\frac{{\gamma ^{\prime 2}}}{4}\int _0^t\int (v_2-v_{*2})^2n_2^2B{\chi}_jff_*F_j(f')F_j(f'_*)dvdv_*d\theta dxds\\
\leq &\frac{c_0}{{\gamma ^{\prime 2}}}{(1+t)}+\frac{{\gamma ^{\prime 2}}}{4}\int _0^t\int (v_2-v_{*2})^2n_2^2B{\chi}_jff_*F_j(f')F_j(f'_*)dvdv_*d\theta dxds.
\end{aligned}\]
%Treating the term $2\int _0^t\int p_1(p_3-p_{*3})n_1n_2\bar{\chi}_jBff_*F(f')F(f'_*)dxdpdp_*d\omega ds$ in an analogous way,
It follows that
\begin{eqnarray*}
\int _0^t \int |v-v_*|^2B{\chi}_jff_*F_j(f')F_j(f'_*)dvdv_*d\theta dxds\leq c'_0(1+t),
\end{eqnarray*}
with $c'_0$ only depending on $\int f_0(x,v)dxdv$ and $\int \lvert v\rvert ^2f_0(x,v)dxdv$. This completes the proof of the lemma. \cqfd
%
%
% Lemma 3.3
%
\[\]
\begin{lemma}
\hspace*{0.2in}\\
{There exist  constants $c'_1$ and $c'_2$ only depending on 
%$\int (1+|v|^2)f_0(x,v)dxdv$, 
$\int f_0(x,v)dxdv$ and on $\int |v|^2f_0(x,v)dxdv$, so that
\begin{eqnarray*}
\int \sup_{0\leq t\leq T}f_j^\sharp (t,x,v)dxdv<c'_1+c'_2T,\quad j\in \N ^*\quad T>0.
\end{eqnarray*}
}
\end{lemma}
\underline{Proof of Lemma 3.3.} \\
Denote $f_j$ by $f$ for simplicity. Since
\begin{eqnarray*}
 f^\sharp (t,x,v)=f_0(x,v)+\int_0^tQ_j(f)(s,x+sv_1,v)ds,
 \end{eqnarray*}
 it holds that
\begin{equation}\label{control-by-gain}
{\sup_{0\leq t\leq T}f^\sharp (t,x,v)\leq  f_0(x,v)+\int_0^{T }Q_j^+(f)(t,x+tv_1,v)dt.}
\end{equation}
Integrating (3.7) with respect to $(x,v)$ and using Lemma 3.2, gives
{\[ \begin{aligned}
\int \sup_{0\leq t\leq T}f^\sharp (t,x,v)dxdv
&\leq  \int f_0(x,v)dxdv+\frac{1}{\pi }\int_0^{T }\int B{\chi}_j\\
f(t,x+tv_1,v')f(t,x+tv_1,v'_*)&F_j(f)(t,x+tv_1,v)F_j(f)(t,x+tv_1,v_*)dvdv_*d\theta dxdt\\
&\leq  \int f_0(x,v)dxdv+\frac{1}{\gamma^2}\int_0^{T }\int B{\chi}_j|v-v_*|^2
\\
f(t,x,v')f(t,x,v'_*)&F_j(f)(t,x,v)F_j(f)(t,x,v_*)dvdv_*d\theta dxdt\\
&\leq  \int f_0(x,v)dxdv+\frac{C_1+C_2T}{\gamma^2}
.\quad \quad \quad \quad \quad \cqfd
\end{aligned}\]
}
%
%
% Lemma 3.4
\[\]
\begin{lemma}
\hspace*{0.2in}\\
Given $T>0$ and $\delta_1>0$, there exist $\delta_2>0$ and $t_0>0$, only depending on 
%$\int (1+|v|^2)f_0(x,v)dxdv$,
$\int f_0(x,v)dxdv$ and on $\int |v|^2f_0(x,v)dxdv$, {such} that for $t\leq T$
\begin{eqnarray*}
\sup _{x_0\in[0,1] }\int_{|x-x_0|<\delta_2} \hspace*{0.03in}{\sup_{t\leq s\leq t+t_0}}f_j^\sharp (s,x,v)dxdv<\delta_1,\quad j\in \N ^*.
\end{eqnarray*}
\end{lemma}
\underline{Proof of Lemma 3.4.} \\
Denote $f_j$ by $f$ for simplicity. For $t\leq s\leq t+t_0$ it holds,
\[ \begin{aligned}
 f^\sharp (s,x,v)&=f^{\sharp}(t+t_0,x,v)-\int_{s}^{t+t_0}Q_j(f)(\tau,x+\tau v_1,v)d\tau\\
&\leq  f^{\sharp}(t+t_0,x,v)+\int_{s}^{t+t_0}Q_j^-(f)(\tau,x+\tau v_1,v)d\tau.
\end{aligned}\]
And so
\begin{eqnarray*}
 \sup_{t\leq s\leq t+t_{0}}f^\sharp (s,x,v)
\leq  f^{\sharp}(t+t_0,x,v)+\int_{t}^{t+t_0}Q_j^-(f)(s,x+sv_1,v)ds.
\end{eqnarray*}
Integrating with respect to $(x,v)${, using Lemma 3.2 and the bound $\frac{1}{\alpha }$ from above of $f$}, gives
\[ \begin{aligned}
&\int_{|x-x_0|<\delta_2} \sup_{t\leq s\leq t+t_0}f^\sharp (s,x,v)dxdv\\
&\leq  \int_{|x-x_0|<\delta_2} f^{\sharp}(t+t_0,x,v)dxdv\\
&+\frac{1}{\pi }\int_{t}^{t+t_0}\int B{\chi}_jf^{\sharp}(s,x,v)f(s,x+sv_1,v_*)F_j(f)(s,x+sv_1,v')
F_j(f)(s,x+sv_1,v'_*)dvdv_*d\theta dxds\\
&\leq  \int_{|x-x_0|<\delta_2} f^{\sharp}(t+t_0,x,v)dxdv
+\frac{1}{\lambda^2}\int_{t}^{t+t_0}\int_{|v-v_*|\geq\lambda} B{\chi}_j|v-v_*|^2
f^{\sharp}(s,x,v)f(s,x+sv_1,v_*)\\
&F_j(f)(s,x+sv_1,v')F_j(f)(s,x+sv_1,v'_*)dvdv_*d\theta dxds\\
&+c\int_{t}^{t+t_0}\int _{|v-v_*|<\lambda} B{\chi}_j
f^{\sharp}(s,x,v)f(s,x+sv_1,v_*)dvdv_*d\theta dxds\\
&\leq  \int_{|x-x_0|<\delta_2} f^{\sharp}(t+t_0,x,v)dxdv
+\frac{C_1+C_2T}{\lambda^2}+ct_0 \lambda^2\int f^\sharp (t,x,v)dxdv\\
&\leq \frac{1}{\Lambda^2}\int v^2f_0dxdv + c\delta_2 \Lambda^2+\frac{C_1+C_2T}{\lambda^2}+ct_0 \lambda^2\int f_0 (x,v)dxdv
\end{aligned}\]
Depending on $\delta_1$, suitably choosing $\Lambda$ and then $\delta_2$, $\lambda$ and then $t_0$, the lemma follows.   \cqfd
The previous lemmas imply {a $t$-dependent bound} for the $v$-integral of $f_j^\#$ only depending on  
%$\int (1+|v|^2)f_0(x,v)dxdv$, 
$\int f_0(x,v)dxdv$ and on $\int |v|^2f_0(x,v)dxdv$, as will now be proved.
%
%
% Lemma 3.5
%
\begin{lemma}
\hspace*{0.1in}\\
Given $T>0$, the solution $f_j$ of (\ref{eq-f_j}) satisfies
\begin{eqnarray*}
\int \sup_{(t,x)\in [0,T] \times [ 0,1] }f^\sharp _j(t,x,v)dv\leq c_1(T),\quad j\in \N ^*,
\end{eqnarray*}
where $c_1(T)$ only depends on $T$, $\int f_0(x,v)dxdv$ and $\int |v|^2f_0(x,v)dxdv$.
%$\int (1+|v|^2)f_0(x,v)dxdv$.
\end{lemma}
\underline{Proof of Lemma 3.5.} \\
For any $a,b\in \R $, denote by $I(a,b)$ the interval with end points $a$ and $b$. \\
Denote by $E(x)$ the integer part of $x\in\R$, $E(x)\leq x< E(x)+1$.\\
As in the proof of Lemma 3.3,
\begin{eqnarray}
 \sup_{s\leq t}f^\sharp (s,x,v)
\leq  f_0(x,v)+\int_0^tQ_j^+(f)(s,x+sv_1,v)ds= f_0(x,v)\hspace*{2.4in}\nonumber \\
+\int_0^t\int B{\chi}_j f(s,x+sv_1,v')f(s,x+sv_1,v'_*)F_j(f)(s,x+sv_1,v)F_j(f)(s,x+sv_1,v_*)dv_*d\omega ds\nonumber\\
\leq  f_0(x,v)+cA,\hspace*{4.5in}
\end{eqnarray}
where
\begin{eqnarray*}
A= \int_0^t\int B{\chi}_j
 \sup _{\tau \in [ 0,t]} f^{\#}(\tau ,x+s(v_1-v^\prime _1),v^\prime ) \sup _{\tau \in [ 0,t]}f^{\#}(\tau ,x+s(v_1-{v^\prime }_{*1}),v^\prime _*)dv_*d\omega ds.\quad
\end{eqnarray*}
For $\theta $ outside of the angular cutoff {(2.2)}, let $n$ be the unit vector in the direction $v-v'$, and $n_{\perp}$ the orthogonal
unit vector in the direction $v-v'_*$. With $e_1$ a unit vector in the $x$-direction,
\begin{eqnarray*}
\max(|n\cdot e_1|,|n_{\perp}\cdot e_1|)\geq \frac{1}{\sqrt{2}}.
\end{eqnarray*}
For $\delta _2>0$ that will be fixed later, split $A$ into $A_1+A_2+A_3+A_4$, where
\begin{eqnarray*}
A_1= \int_0^t\int _{|n\cdot e_1|\geq \frac{1}{\sqrt{2}},\hspace*{0.02in}t|v_1-v^\prime _1|>\delta _2}B{\chi}_j\sup _{\tau \in [ 0,t]} f^{\#}(\tau ,x+s(v_1-v'_1),v') \sup _{\tau \in [ 0,t]}f^{\#}(\tau ,x+s(v_1-{v'}_{*1}),v'_*)dv_*d\omega ds,\\
A_2= \int_0^t\int _{|n\cdot e_1|\geq \frac{1}{\sqrt{2}},\hspace*{0.02in}t|v_1-v^\prime _1|<\delta _2}B{\chi}_j\sup _{\tau \in [ 0,t]} f^{\#}(\tau ,x+s(v_1-v'_1),v') \sup _{\tau \in [ 0,t]}f^{\#}(\tau ,x+s(v_1-{v'}_{*1}),v'_*)dv_*d\omega ds,\\
A_3= \int_0^t\int _{|n_\perp \cdot e_1|\geq \frac{1}{\sqrt{2}},\hspace*{0.02in}t|v_1-v^\prime _1|>\delta _2}B{\chi}_j\sup _{\tau \in [ 0,t]} f^{\#}(\tau ,x+s(v_1-v'_1),v') \sup _{\tau \in [ 0,t]}f^{\#}(\tau ,x+s(v_1-{v'}_{*1}),v'_*)dv_*d\omega ds,\\
A_4= \int_0^t\int _{|n_\perp \cdot e_1|\geq \frac{1}{\sqrt{2}},\hspace*{0.02in}t|v_1-v^\prime _1|<\delta _2}B{\chi}_j\sup _{\tau \in [ 0,t]} f^{\#}(\tau ,x+s(v_1-v'_1),v') \sup _{\tau \in [ 0,t]}f^{\#}(\tau ,x+s(v_1-{v'}_{*1}),v'_*)dv_*d\omega ds.
\end{eqnarray*}
In $A_1$ and $A_2$, bound the factor $\sup _{\tau \in [ 0,t] }f^\sharp (\tau ,x+s(v_1-v^\prime _{*1}),v^\prime _*)$ by its supremum over $x\in [ 0,1] $, and make the change of variables
\begin{eqnarray*}
 s\rightarrow y= x+s(v_1-v^\prime _1).
\end{eqnarray*}
with Jacobian
\begin{eqnarray*}
\frac{Ds}{Dy}= \frac{1}{|v_1-v^\prime _1|}= \frac{1}{|v-v_*|\hspace*{0.03in}|(n,\frac{v-v_*}{|v-v_*|})|\hspace*{0.03in}|n_1|} \leq \frac{\sqrt{2}}{\gamma \gamma ^\prime }.
\end{eqnarray*}
It holds that
\begin{eqnarray*}
A_1\leq \int _{t|v_1-v^\prime _1|>\delta _2}\frac{B{\chi}_j}{|v_1-v^\prime _1|}\Big( \int _{y\in I(x,x+t(v_1-v^\prime _1)}\sup _{\tau \in [ 0,t]} f^{\#}(\tau ,y,v')dy\Big)  \sup _{(\tau ,X)\in [ 0,t]\times [ 0,1] }f^{\#}(\tau ,X,v'_*)dv_*d\omega ,
\end{eqnarray*}
and
\begin{eqnarray*}
A_2\leq \frac{\sqrt{2}}{\gamma \gamma ^\prime }\int _{|n\cdot e_1|\geq \frac{1}{\sqrt{2}},\hspace*{0.02in}t|v_1-v^\prime _1|<\delta _2}B{\chi}_j\Big( \int _{|y-x|<\delta _2}\sup _{\tau \in [ 0,t]} f^{\#}(\tau ,y,v')dy\Big)  \sup _{(\tau ,X)\in [ 0,t]\times [ 0,1] }f^{\#}(\tau ,X,v'_*)dv_*d\omega .
\end{eqnarray*}
Then, performing the change of variables $(v,v_*,\omega )\rightarrow (v^\prime ,v^\prime _*,-\omega )$,
\[ \begin{aligned}
&\int \sup _{x\in [ 0,1] }A_1dv\\
&\leq \int _{t|v_1-v^\prime _1|>\delta _2}\frac{B{\chi}_j}{|v_1-v^\prime _1|}\sup _{x\in [ 0,1] }\Big( \int _{y\in I(x,x+t(v^\prime _1-v_1)}\sup _{\tau \in [ 0,t]} f^{\#}(\tau ,y,v)dy\Big)  \sup _{(\tau ,X)\in [ 0,t]\times [ 0,1] }f^{\#}(\tau ,X,v_*)dvdv_*d\omega ,
\end{aligned}\]
so that
\[ \begin{aligned}
&\int \sup _{x\in [ 0,1] }A_1dv\\
&\leq \int _{t|v_1-v^\prime _1|>\delta _2}\frac{B{\chi}_j}{|v_1-v^\prime _1|}\sup _{x\in [ 0,1] }\Big( \int _{y\in I(x,x+E(t(v^\prime _1-v_1)+1)}\sup _{\tau \in [ 0,t]} f^{\#}(\tau ,y,v)dy\Big)  \sup _{(\tau ,X)\in [ 0,t]\times [ 0,1] }f^{\#}(\tau ,X,v_*)dvdv_*d\omega \\
&= \int _{t|v_1-v^\prime _1|>\delta _2}\frac{B{\chi}_j}{|v_1-v^\prime _1|} | E(t(v^\prime _1-v_1)+1)|\Big( \int _0^1\sup _{\tau \in [ 0,t]} f^{\#}(\tau ,y,v)dy\Big)  \sup _{(\tau ,X)\in [ 0,t]\times [ 0,1] }f^{\#}(\tau ,X,v_*)dvdv_*d\omega \\
&\leq t(1+\frac{1}{\delta _2})\int B{\chi}_j\Big( \int _0^1\sup _{\tau \in [ 0,t]} f^{\#}(\tau ,y,v)dy\Big)  \sup _{(\tau ,X)\in [ 0,t]\times [ 0,1] }f^{\#}(\tau ,X,v_*)dvdv_*d\omega \\
&\leq B_0\textcolor{red}{\pi }t(1+\frac{1}{\delta_2})\int \sup _{\tau \in [ 0,t]} f^{\#}(\tau ,y,v)dydv\int \sup _{(\tau ,X)\in [ 0,t]\times [ 0,1] }f^{\#}(\tau ,X,v_*)dv_*.
\end{aligned}\]
Apply Lemma 3.3, so that
\begin{equation}\label{bdd-A1}
\int \sup _{x\in [ 0,1] }A_1dv\leq (c_1^\prime+c'_2T) B_0\pi  t(1+\frac{1}{\delta_2})\int \sup _{(\tau ,X)\in [ 0,t]\times [ 0,1] }f^{\#}(\tau ,X,v_*)dv_*.
\end{equation}
Moreover, performing the change of variables $(v,v_*,\omega )\rightarrow (v^\prime _*,v^\prime ,-\omega )$,
\begin{eqnarray*}
\int \sup _{x\in [ 0,1] }A_2dv\leq \frac{B_0\pi \sqrt{2}}{\gamma \gamma ^\prime }\sup _{x\in [ 0,1] }\Big( \int _{|y-x|<\delta _2}\sup _{\tau \in [ 0,t]} f^{\#}(\tau ,y,v_*)dydv_*\Big)  \int \sup _{(\tau ,X)\in [ 0,t]\times [ 0,1] }f^{\#}(\tau ,X,v)dv.
\end{eqnarray*}
Given $\delta_1= \frac{\gamma \gamma ^\prime }{4B_0\pi \sqrt{2}}$, apply Lemma 3.4 with the corresponding $\delta_2$ and $t_0$, so that for $t\leq t_0$,
\begin{equation}\label{bdd-A2}
\int \sup _{x\in [ 0,1] }A_2dv\leq \frac{1}{4}\int  \sup _{(\tau ,X)\in [ 0,\textcolor{blue}{t}]\times [ 0,1] }f^{\#}(\tau ,X,v)dv.
\end{equation}
The terms $A_3$ and $A_4$ are treated similarly, with the change of variables $ s\rightarrow y= x+s(v_1-v'_{*1})$. \\
Using (\ref{bdd-A1})-(\ref{bdd-A2}) and the corresponding bounds obtained for $A_3$ and $A_4$ leads to
\[ \begin{aligned}
\int \sup_{(s,x)\in [ 0,t ]\times [0,1]}f^{\#}(s,x,v)dv\leq &2\int \sup_{x\in [ 0,1] } f_0(x,v)dv\\
+{4}(c^\prime _1+c'_2T)B_0\pi  t (1+\frac{1}{\delta_2})\int\sup_{(s,x)\in [ 0,t] \times [0,1]} f^{\#}(s,x,v)dv,\quad t\leq t_0.
\end{aligned}\]
Hence for $t\leq\min(t_0, (8(c^\prime _1+c'_2T)B_0\pi ^2  (1+\frac{1}{\delta_2}))^{-1}$
\begin{eqnarray*}
\int \sup_{(s,x)\in [ 0,t ]\times [0,1]}f^{\#}(s,x,v)dv\leq 4\int \sup_{x\in [ 0,1] } f_0(x,v)dv.
\end{eqnarray*}
Since $c'_1$, $c'_2$, and $t_0$ only depend on $\int (1+|v|^2)f_0(x,v)dxdv$ and $T$, it follows that the argument can be repeated up to $t=T$. This completes the proof of the lemma.                       \cqfd

\hspace*{0.1in}\\
\hspace*{0.1in}\\
\section{Proof of the main theorem and the asymptotic behavior.}
\setcounter{theorem}{0}
\setcounter{equation}{0}
The following  two preliminary lemmas are needed for the control of large velocities.
\begin{lemma}
\hspace*{0.02in}\\
Given $t>0$, there is a constant $c_t>0$ such that the solutions of (\ref{eq-f_j}) satisfy
\begin{eqnarray*}
\sup _{j\in \N ^* } \int _0^1\int_{|v|>\lambda} |v|\sup_{s\leq t}f_j^\sharp (s,x,v)dvdx\leq\frac{c_t}{\lambda}.
\end{eqnarray*}
\end{lemma}
\underline{Proof of Lemma 4.1.} \\
For convenience $j$ is dropped from the notation $f_j$. As in Section 3,
\begin{eqnarray*}
 \sup_{s\leq t}f^\sharp (s,x,v)
\leq  f_0(x,v)+\int_0^tQ_j^+(f)(s,x+sv_1,v)ds.
\end{eqnarray*}
Integration with respect to $(x,v)$  for $|v|>\lambda$, gives
\begin{eqnarray*}
 \int _0^1\int_{|v|>\lambda}|v|\sup_{s\leq t}f^\sharp (s,x,v)dvdx
\leq  \int \int_{|v|>\lambda}|v|f_0(x,v)dvdx+\int_0^t\int_{|v|>\lambda} B{\chi}_j\\
|v| f(s,x+sv_1,v^\prime )f(s,x+sv_1,v^\prime _*)F(f)(s,x+sv_1,v)F(f)(s,x+sv_1,v_*)dvdv_*d\omega dxds.
\end{eqnarray*}
Here in the last integral, either $|v^\prime |$ or $|v^\prime _*|$ is the largest and larger than $\frac{\lambda}{\sqrt 2}$. The two cases are symmetric, and we discuss the case $|v^\prime |\geq|v^\prime _*|$. After a translation in $x$, the integrand is estimated {from above} by
$c |v^\prime |f^{\#} (s,x,v^\prime )\sup_{x\in [0,1], s\leq t}f^{\#}(s,x,v^\prime _*)$. The change of variables $(v,v_*,\omega )\rightarrow (v^\prime ,v^\prime _*,-\omega )$, the integration over $(s,x,v,v_*,\omega )\in [ 0,t] \times [ 0,1] \times \{ v\in \R ^2; |v| >\frac{\lambda}{\sqrt 2}\} \times \R ^2 \times S^1$ and Lemma 3.5 give the bound
\begin{eqnarray*}
\frac{c}{\lambda}\Big( \int_0^t\int |v|^2f^{\#}(s,x,v)dxdvds\Big) \Big( \int \sup_{s\leq t,x\in [0,1]} f^{\#}(s,x,v_*)dv_*\Big) \leq \frac{ctc_1(t)}{\lambda}\int |v|^2f_0(x,v)dxdv.
\end{eqnarray*}
The lemma follows.                        \cqfd
%
%
% Lemma 4.2
%
\begin{lemma}
\hspace*{0.02in}\\
 Given $t>0$ {and $\lambda>2$}, there is a constant $c_t^\prime >0$, such that the solutions $f_j$ of (\ref{eq-f_j}) satisfy
\begin{eqnarray*}
\sup _{j\in \N ^* }\int_{|v|>\lambda} \sup_{(s,x)\in [ 0,t] \times [0,1]}f_j^\sharp (s,x,v)dv\leq\frac{c'_t}{\sqrt{\lambda}}.
\end{eqnarray*}
\end{lemma}
\underline{Proof of Lemma 4.2.}\\
Take $\lambda >2$. As above,
\begin{eqnarray}
 \int _{|v|>\lambda }\sup_{(s,x)\in [ 0,t] \times [ 0,1] }f^\sharp (s,x,v)dv
\leq  \int _{|v|>\lambda }\sup_{x\in [ 0,1] }f_0(x,v)dv+C,
\end{eqnarray}
where
\begin{eqnarray*}
C= c\int _{|v|>\lambda }\sup _{x\in [ 0,1] }\int_0^t\int B{\chi}_j
 f^{\#}(s,x+s(v_1-v'_1),v')f^{\#}(s,x+s(v_1-v'_{*1}),v'_*)dvdv_*d\omega ds.
\end{eqnarray*}
For $v',v'_*$ outside of the angular cutoff {(2.2)}, let $n$ be the unit vector in the direction $v-v'$, and $n_{\perp}$ the orthogonal
unit vector in the direction $v-v'_*$. Let $e_1$ be a unit vector in the $x$-direction.\\
Split $C$ as $C= \sum _{1\leq i\leq 6}C_i$, where $C_1$ (resp. $C_2$, $C_3$) refers to integration on
\begin{eqnarray*}
\{ (v_*,\omega ); \quad n\cdot e_1\geq \frac{1}{\sqrt{2}}, \quad  |v'| \geq |v'_*|\} ,
\end{eqnarray*}
\begin{eqnarray*}
\big( \text{resp.    }\{ (v_*,\omega ); n\cdot e_1\geq \sqrt{1-\frac{1}{\lambda }}, \hspace*{0.03in}|v'| \leq |v'_*|\} ,\quad \{ (v_*,\omega ); n\cdot e_1\in [\frac{1}{\sqrt{2}}, \sqrt{1-\frac{1}{\lambda }}], \hspace*{0.03in}|v'| \leq |v'_*|\}\big) ,
\end{eqnarray*}
and analogously for $C_i$, $4\leq i\leq 6$, with $n$ replaced by $n_\perp $.
By symmetry, $C_i$, $4\leq i\leq 6$ can be treated as $C_i$, $1\leq i\leq 3$, so we only discuss the control of $C_i$, $1\leq i\leq 3$.\\
By the change of variables $(v,v_*,\omega)\rightarrow (v^\prime ,v^\prime _*,-\omega )$, and noticing that $ |v'| \geq \frac{\lambda }{\sqrt{2}}$ in the domain of integration of $C_1$, it holds that
\[ \begin{aligned}
C_1&\leq \int _{|v|>\frac{\lambda }{\sqrt{2}}}\sup _{x\in [ 0,1] }\int_0^t\int _{n\cdot e_1\geq \frac{1}{\sqrt{2}}}B{\chi}_j
 f^{\#}(s,x+s(v^\prime _1-v_1),v)f^{\#}(s,x+s(v^\prime _1-v_{*1}),v_*)dv_*d\omega dsdv\\
& \leq \int _{|v|>\frac{\lambda }{\sqrt{2}}}\sup _{x\in [ 0,1] }\int_0^t\int _{n\cdot e_1\geq \frac{1}{\sqrt{2}}}B{\chi}_j
 \sup _{\tau \in [ 0,t] }f^{\#}(\tau ,x+s(v^\prime _1-v_1),v)\sup _{(\tau ,X)\in [ 0,t] \times [ 0,1] }f^{\#}(\tau ,X,v_*)dv_*d\omega dsdv.
\end{aligned}\]
With the change of variables $s\rightarrow y= x+s(v^\prime _1-v_1)$,
\[ \begin{aligned}
C_1&\leq \int _{|v|>\frac{\lambda }{\sqrt{2}}}\sup _{x\in [ 0,1] }\int _{n\cdot e_1\geq \frac{1}{\sqrt{2}}}\int _{y\in I(x,x+t(v^\prime _1-v_1))}\frac{B{\chi}_j}{|v^\prime _1-v_1|}
  \sup _{\tau \in [ 0,t] }f^{\#}(\tau ,y,v)\sup _{(\tau ,X)\in [ 0,t] \times [ 0,1] }f^{\#}(\tau ,X,v_*)dydv_*d\omega dv\\
   &\leq  \int _{|v|>\frac{\lambda }{\sqrt{2}}}\int _{n\cdot e_1\geq \frac{1}{\sqrt{2}}} \frac{{ |E(t(v^\prime _1-v_1))+1) |}}{|v^\prime _1-v_1|}\int _0^1B{\chi}_j
  \sup _{\tau \in [ 0,t] }f^{\#}(\tau ,y,v)\sup _{(\tau ,X)\in [ 0,t] \times [ 0,1] }f^{\#}(\tau ,X,v_*)dydv_*d\omega dv.
  \end{aligned}\]
  {Moreover,
  \begin{eqnarray*}
  |E(t(v^\prime _1-v_1))+1) | \leq t | v_1^\prime -v_1| +{1}\leq \big( t+\frac{{\sqrt{2}}}{\gamma \gamma ^\prime }\big) | v_1^\prime -v_1| .
  \end{eqnarray*}
  Consequently, }
  \[ \begin{aligned}
  C_1&\leq  c(t+1)\int _0^1\int _{|v|>\frac{\lambda }{\sqrt{2}}}
  \sup _{\tau \in [ 0,t] }f^{\#}(\tau ,y,v)dydv\int \sup _{(\tau ,X)\in [ 0,t] \times [ 0,1] }f^{\#}(\tau ,X,v_*)dv_*\\
 &\leq  \frac{c(t+1)}{\lambda }\int _0^1\int _{|v|>\frac{\lambda }{\sqrt{2}}}
  |v|\sup _{\tau \in [ 0,t] }f^{\#}(\tau ,y,v)dydv\int \sup _{(\tau ,X)\in [ 0,t] \times [ 0,1] }f^{\#}(\tau ,X,v_*)dv_*.
\end{aligned}\]
By Lemma 3.5 and Lemma 4.1,
\begin{eqnarray*}
C_1\leq\frac{c}{\lambda^2}(t+1)c_t c_1(t).
\end{eqnarray*}
Moreover,
\[ \begin{aligned}
C_2 &\leq \int _{|v^\prime |>\lambda , |v_*|> |v|, n\cdot e_1\geq \sqrt{1-\frac{1}{\lambda }}}\frac{B\chi _j}{|v^\prime _1-v_1|}\\
&\sup _{x\in [ 0,1] }\int_{y\in I(x,x+t(v^\prime _1-v_1)}
 \sup _{\tau \in [ 0,t] }f^{\#}(\tau ,y,v)
\sup _{(\tau ,X)\in [ 0,t] \times [ 0,1] }f^{\#}(\tau ,X,v_*)dydvdv_*d\omega \\
&\leq  c(t+1)\int _{n\cdot e_1\geq \sqrt{1-\frac{1}{\lambda }}}d\omega \int
 \sup _{\tau \in [ 0,t] }f^{\#}(\tau ,y,v)dydv\int \sup _{(\tau ,X)\in [ 0,t] \times [ 0,1] }f^{\#}(\tau ,X,v_*)dv_*\\
& \leq \frac{c}{\sqrt{\lambda }}(t+1)^2c_1(t),
\end{aligned}\]
by Lemma 3.3 and Lemma 3.5.
Finally,
\[ \begin{aligned}
C_3& \leq \int _{|v_*|>\frac{\lambda }{\sqrt{2}}, \frac{1}{\sqrt{\lambda }}\leq n_\perp \cdot e_1\leq \frac{1}{\sqrt{2}}}\sup _{(\tau ,X)\in [ 0,t] \times [ 0,1] }f^{\#}(\tau ,X,v)\frac{B\chi _j}{|v^\prime _1-v_{*1}|}\\
&\hspace*{0.3in}\sup _{x\in [ 0,1] }\Big( \int_{y\in I(x,x+t(v^\prime _1-v_{*1})}
 \sup _{\tau \in [ 0,t] }f^{\#}(\tau ,y,v_*)dy\Big) dvdv_*d\omega \\
&\leq  c(t+1)\sqrt{\lambda}\Big( \int \sup _{(\tau ,X)\in [ 0,t] \times [ 0,1] }f^{\#}(\tau ,X,v)dv\Big) \Big( \int_{|v_*|>\frac{\lambda }{\sqrt{2}}} \sup _{\tau \in [ 0,t] }f^{\#}(\tau ,y,v_*)dydv_*\Big) .
\end{aligned}\]
By Lemma 3.5,
\[ \begin{aligned}
C_3&\leq  c(t+1)\sqrt{\lambda}c_1(t)\int_{|v_*|>\frac{\lambda }{\sqrt{2}}} \sup _{\tau \in [ 0,t] }f^{\#}(\tau ,y,v_*)dydv_*,
\end{aligned}\]
and so by Lemma 4.1,
\[ \begin{aligned}
C_3&\leq  {\frac{c}{\sqrt{\lambda}}(t+1)c_1(t)c_t.}
\end{aligned}\]
\hspace*{0.1in}\\
The lemma follows. \cqfd
%
%
%
% Proof of Theorem 2.1.
%
Using the previous lemmas, the results in Section 3, and an initial layer analysis, the main result of the paper follows.\\
\hspace*{0.1in}\\
\hspace*{0.1in}\\
\underline{Proof of Theorem 2.1.}\\
For any $T>0$, we shall prove the convergence in $C([ 0,T]; L^1([ 0,T] \times \R ^2))$ of the sequence $(f_j)$ to a solution $f$ of (\ref{f}). Denote by
\begin{eqnarray*}
\tilde{\nu}_j(f):=\int B\chi _j f^\prime f^\prime _*F_j(f_*) dv_*d\omega ,\quad \nu _j( f):=\int B\chi _j f_*F_j(f^\prime )F_j(f^\prime _*) dv_*d\omega ,
\end{eqnarray*}
so that
\begin{eqnarray*}
Q_j(f)= F_j(f)\tilde{\nu }_j(f)-f\nu _j(f) .
\end{eqnarray*}
Consider
\[ \begin{aligned}
&\nu _j(f_j)^\sharp (t,x,v)= \int B\chi_jf_j(t,x+tv_1,v_*)F_j(f_j(t,x+tv_1,v^\prime ))F_j(f_j(t,x+tv_1,v^\prime _*))dv_*d\omega.
\end{aligned}\]
{With the angular cut-off (2.2), $v_*  \rightarrow v^\prime $ and $v_*  \rightarrow v^\prime _*$ are changes of variables. Indeed, if the polar coordinates of $v_*-v$ are $(r_*,\varphi)$ and $\theta $ is the angle between $v_*-v$ and $n $, then the polar coordinates of $v^\prime -v$ (resp.  $v^\prime _*-v$) are $(|r_*cos\theta| ,\varphi+\theta )$ (resp. $(|r_*sin\theta| ,\varphi +\theta +\frac{\pi }{2})$). It follows from the angular cut-off (2.2), that the Jacobians $\frac{Dv_*}{Dv^\prime }= \frac{1}{\mid cos\theta \mid }$ (resp. $\frac{Dv_*}{Dv_*^\prime }= \frac{1}{\mid sin\theta \mid }$ are bounded. Using these changes of variables and }
%
% Df of the sets Z
%
 Lemma 3.5, for $\omega$ outside the integration cut-off, the measure of the set
\begin{equation}\label{df-Z-j0}
Z_{(j,t,x,v,\omega )}:= \{v_*;f(t,x+tv_1,v')>\frac{1}{2}\quad \text{or}\quad f(t,{x}+t{v}_1,v'_*)>\frac{1}{2}\}
\end{equation}
is uniformly bounded with respect to $(x,v,\omega)$, $t\leq T$, and $j\in \N ^*$. Take $j_T$ so large that $\pi j_T^2$ is at least eight times this uniform bound. Notice that here  $j_T$ only depends on $T$ and $\int(1+v^2)f_0(x,v)dxdv$. Using the exponential form for the solution, one gets using Lemma 3.5 that
\begin{equation}\label{bdd-below-fj}
f_j^\sharp (t,x,v_*)\geq c_{1T} f_0(x,v_*)>0,\quad j\geq j_T,\quad t\leq T,
\end{equation}
with $c_{1T}$ independent of $j\geq j_T$.
%
%This is the essential estimate to make $j_T$ (and later $\eta_T$) independent of $j$ and $t\leq T$.
%
It follows {from (\ref{bdd-below-fj}) and the third assumption in (2.4)} that
\begin{equation}\label{bdd-below-nu}
\nu _j(f_j)^\sharp (t,x,v)>c_{2T}>0,\quad (t,x,v)\in [ 0,T] \times [ 0,1] \times \{ v\in \R ^2;|v|\leq j\} ,
\end{equation}
uniformly with respect to $j\geq j_T$, and with $c_{2T}$ only depending on $T$ and $f_0$.\\
{Using again the $v_*\rightarrow v^\prime $ change of variables together with Lemma 3.5}, one obtains that for some constant $c_{3T}>0$,
\begin{eqnarray*}
\tilde{\nu }_j^\sharp(f_j) (t,x,v)\leq c_{3T},\quad j\geq j_T,\quad (t,x,v)\in [ 0,T] \times [ 0,1] \times \{ v\in \R ^2;|v|\leq j\} .
\end{eqnarray*}
The functions defined on $] 0,\frac{1}{\alpha }] $ by $x\rightarrow \frac{F_j(x)}{x}$ are uniformly bounded {from above} with respect to $j$ by
\begin{eqnarray*}
x\rightarrow {c}\alpha ^{\alpha -1}\frac{(1-\alpha x)^\alpha }{x},
\end{eqnarray*}
that is continuous and decreasing to zero at $x= \frac{1}{\alpha }$. Hence there is $\mu \in ] 0,\frac{1}{\alpha }[$ such that
\begin{eqnarray*}
x\in{ [ } \frac{1}{\alpha }-\mu ,\frac{1}{\alpha }] \quad \text{implies}\quad \frac{F_j(x)}{x}{\leq \frac{c_{2T}}{4c_{3T}}},\quad j\geq j_T.
\end{eqnarray*}
Consequently, for $j\geq j_T$,
\[ \begin{aligned}
f_j^\sharp (t,x,v)\in { [ } \frac{1}{\alpha }-\mu ,\frac{1}{\alpha }] \quad \Rightarrow \quad D_tf_j^\sharp (t,x,v)&= \big( F_j(f_j^\sharp )\tilde{\nu }_j^\sharp -\frac{1}{2}f_j^\sharp \nu _j^\sharp \big) (t,x,v)-\frac{1}{2}f_j^\sharp \nu _j^\sharp (t,x,v)\\
&< -\frac{1}{2}f_j^\sharp \nu _j^\sharp (t,x,v)\\
&<-\frac{1}{2}(\frac{1}{\alpha }-{\mu} )c_{2T}:=-b_1.
\end{aligned}\]
This gives a maximum time $t_1=\frac{{\mu }}{b_1}$ for $f_j^\#$ to reach $\frac{1}{\alpha}-\mu$ from an initial value $f_0(x,v)\in ] \frac{1}{\alpha }-\mu ,\frac{1}{\alpha }]$.  On this time interval $D_tf_j^\sharp \leq -b_1$. If $t_1\geq T$, then at $t= T$ the value of $f_j^\#$ is bounded from above by $\frac{1}{\alpha}-b_1T:= \frac{1}{\alpha}-\mu ^\prime $ with $0<\mu^\prime \leq \mu$. Take $t_m=\min (t_1,T)$, and from now on $\mu= t_mb_1$. For any  $(x,v)$, if $f_j(0,x,v)< \frac{1}{\alpha }-\mu $ were to reach $\frac{1}{\alpha}-\mu$ at $(t,x,v)$ with $t\leq t_m$, then $D_tf_j^\#(t,x,v)\leq -b_1$, which excludes such a possibility.
%If $t_1>t_0$, then at $t=t_0$ for some $b_2$ with $0<b_2\leq b_1$, analogously
It follows that $f_j\leq \frac{1}{\alpha}-\mu$ everywhere for $t\in [ t_m, T] $, and {that\\
%\textcolor{red}{Could you especially check the following (4.5):
\begin{equation}\label{interval0-tm}
%f_j^\sharp (t,x,v)\in [ \frac{1}{\alpha }-\mu ,\frac{1}{\alpha }] \Rightarrow
 f_j^\sharp (t,x,v)\leq \frac{1}{\alpha }-b_1t.
\end{equation}
for $t\in[0,t_m]$. The previous estimates leading to the definition of $t_m$ are independent of $j\geq j_T$.\\
\hspace*{0.1in}\\
Let us prove that  $(f_j)$ converges {in $L^1( [ 0,T] \times [ 0,1] \times \R ^2)$} when $j\rightarrow \infty$. \\
We shall prove that given $\beta>0$, there exists $a\geq \max \{ 1, j_T\} $, so that
\begin{equation}
\sup_{{t\in }[0,T]}\int |g_j(t,x,v)|dxdv< \beta,\quad j>a,
\end{equation}
where $g_j=f_j-f_a$. The function $g_j$ satisfies the equation
\[ \begin{aligned}
\partial _tg_j+p_1\partial _xg_j&= \int ({\chi} _j-{\chi}_a)B\Big( f_j^\prime f^\prime _{j*}F_j(f_j)F_j(f_{j*})- f_jf_{j*}F_j(f^\prime _j)F_j(f^\prime _{j*})\Big) dv_*d\omega\\
&+\int \chi _a B(f_j^\prime f_{j*}^\prime -f_a^\prime f_{a*}^\prime )F_j(f_j)F_j(f_{j*})dv_*d\omega \\
&-\int \chi _a B(f_jf_{j*} -f_af_{a*})F_j(f_j^\prime )F_j(f_{j*}^\prime )dv_*d\omega \\
&+\int \chi _aBf_a^\prime f_{a*}^\prime \Big( F_j(f_{j*})\big( F_j(f_j)-F_j(f_a)\big) +F_a(f_a)\big( F_j(f_{j*})-F_j(f_{a*})\big) \Big) dv_*d\omega \\
&+\int \chi _aBf_a^\prime f_{a*}^\prime \Big( F_j(f_{j*})\big( F_j(f_a)-F_a(f_a)\big) +F_a(f_a)\big( F_j(f_{a*})-F_a(f_{a*})\big) \Big) dv_*d\omega \\
& -\int \chi _aBf_af_{a*}\Big( F_j(f_{j*}^\prime )\big( F_j(f_j^\prime )-F_j(f_a^\prime )\big) +F_a(f_a^\prime )\big( F_j(f_{j*}^\prime )-F_j(f_{a*}^\prime )\big) \Big) dv_*d\omega \\
&-\int \chi _aBf_af_{a*}\Big( F_j(f_{j*}^\prime )\big( F_j(f_a^\prime )-F_a(f_a^\prime )\big) +F_a(f_a^\prime )\big( F_j(f_{a*}^\prime )-F_a(f_{a*}^\prime )\big) \Big) dv_*d\omega .\quad (4.7)
\end{aligned}\]
Moreover, {using Lemma 3.5}
\[ \begin{aligned}
\int ({\chi} _j-{\chi}_a)B&\Big( f_j^\prime f^\prime _{j*}F_j(f_j)F_j(f_{j*})+f_jf_{j*}F_j(f^\prime _j)F_j(f^\prime _{j*})\Big) dxdvdv_*d\omega\\
& \leq c\int _{\lvert v\rvert >\frac{a}{\sqrt{2}}}f_j(t,x,v)dxdv\\
&\leq \frac{c}{a^2}\text{     by the conservation of energy of   }f_j,
\end{aligned}\]
\[ \begin{aligned}
\int \chi _a B&\lvert f_jf_{j*} -f_af_{a*}\rvert F_j(f_j^\prime )F_j(f_{j*}^\prime )dxdvdv_*d\omega \\
&\leq c\Big( \int \sup _{(t,x)\in [ 0,T] \times [ 0,1] }f_j^\sharp (t,x,v)dv
+ \int \sup _{(t,x)\in [ 0,T] \times [ 0,1] }f_a^\sharp (t,x,v)dv\Big) \int \lvert (f_j^\sharp -f_a^\sharp )(t,x,v)\rvert dxdv\\
&\leq c\int \lvert (f_j^\sharp -f_a^\sharp )(t,x,v)\rvert dxdv\quad \text{  by Lemma 3.5.}
\end{aligned}\]
Next,
\[ \begin{aligned}
&\int \chi _aB\Big( f_a^\prime f_{a*}^\prime F_j(f_{j*})\lvert  F_j(f_a)-F_a(f_a)\rvert \Big) ^\sharp dxdvdv_*d\omega \\
&= \int \chi _aBf_a^\prime f_{a*}^\prime F_j(f_{j*})
(1-\alpha f_a) (1+(1-\alpha)f_a)^{1-\alpha} \lvert (\frac{1}{j}+1-\alpha f_a)^{\alpha-1}
-(\frac{1}{a}+1-\alpha f_a)^{\alpha-1}\rvert dxdvdv_*d\omega .
\end{aligned}\]
By Lemma 3.3 and Lemma 3.5, this integral restricted to the set where $1-\alpha f_a(t,x,v))\leq\frac{2}{a}$, {hence where
\begin{eqnarray*}
(1-\alpha f_a)  \lvert (\frac{1}{j}+1-\alpha f_a)^{\alpha-1}
-(\frac{1}{a}+1-\alpha f_a)^{\alpha-1}\rvert \leq 2(1-\alpha f_a)^\alpha \leq \frac{2^{\alpha +1}}{a^\alpha },
\end{eqnarray*}
is bounded by $\frac{c}{a^\alpha}$ for some constant $c>0$. } \\
For the remaining domain of integration where $1-\alpha f_a(t,x,v))\geq\frac{2}{a}$, {it holds
\[ \begin{aligned}
|F_j(f_a)-F_a(f_a)| &\leq c (1-\alpha f_a)^\alpha\lvert (\frac{1}{j(1-\alpha f_a)}+1)^{\alpha-1}
-(\frac{1}{a(1-\alpha f_a)}+1)^{\alpha-1}\rvert \\
&= c(\frac{1}{j}-\frac{1}{a})(1-\alpha f_a)^{\alpha -1}\lambda ^{\alpha -2}\quad \text{where    }\lambda \in [ 1,\frac{3}{2}]\\
&\leq \frac{2^{\alpha -1}c}{a^\alpha}.
\end{aligned}\]
And so,
\[ \begin{aligned}
&\int \chi _aB\Big( f_a^\prime f_{a*}^\prime F_j(f_{j*})\lvert  F_j(f_a)-F_a(f_a)\rvert \Big) ^\sharp dxdvdv_*d\omega \leq \frac{c}{a^\alpha }.
\end{aligned}\]
}
Finally

\begin{eqnarray*}
\int \chi _aB\Big( f_a^\prime f_{a*}^\prime F_j(f_{j*})\lvert F_j(f_j)-F_j(f_a)\rvert \Big) ^\sharp (t,x,v)dxdvdv_*d\omega
\leq  c\int \lvert F_j(f_j)-F_j(f_a)\rvert ^\sharp (t,x,v)dxdv.\hspace{1.5in}
\end{eqnarray*}
Split the $(x,v)$-domain of integration of the latest integral into
\[ \begin{aligned}
&D_1:= \{ (x,v); (f_j^\sharp (t,x,v),f_a^\sharp (t,x,v))\in [ 0,\frac{1}{\alpha }-\mu ] ^2\} ,\\
&D_2:= \{ (x,v); (f_j^\sharp (t,x,v),f_a^\sharp (t,x,v))\in [ \frac{1}{\alpha }-\mu ,\frac{1}{\alpha }] ^2\} ,\\
&D_3:= \{ (x,v); (f_j^\sharp ,f_a^\sharp )(t,x,v)\in [ \frac{1}{\alpha }-\mu ,\frac{1}{\alpha }] \times [ 0,\frac{1}{\alpha }-\mu ] \text{   or   } (f_j^\sharp ,f_a^\sharp )(t,x,v))\in [ 0,\frac{1}{\alpha }-\mu ] \times [ \frac{1}{\alpha }-\mu ,\frac{1}{\alpha } ]\} .
\end{aligned}\]
It holds that
\[ \begin{aligned}
&\int _{D_1}\lvert F_j(f_j)-F_j(f_a)\rvert ^\sharp (t,x,v)dxdv\leq c(\alpha \mu )^{\alpha -1}\int _{D_1}\lvert g_j^\sharp (t,x,v)\rvert dxdv,\\
&\int _{D_2}\lvert F_j(f_j)-F_j(f_a)\rvert ^\sharp (t,x,v)dxdv\leq ct^{\alpha -1}\int _{D_2}\lvert g_j^\sharp (t,x,v)\rvert dxdv,\quad {\text{by   }(\ref{interval0-tm}),}\\
&\int _{D_3}\lvert F_j(f_j)-F_j(f_a)\rvert ^\sharp (t,x,v)dxdv\leq c\big( (\alpha \mu )^{\alpha -1}+t^{\alpha -1}\big) \int _{D_3}\lvert g_j^\sharp (t,x,v)\rvert dxdv.
\end{aligned}\]
The remaining terms to the right in (4.7) are of the same types as the ones just estimated. Consequently,
\[ \begin{aligned}
\frac{d}{dt}\int |g_j^\sharp (t,x,v)|dxdv&\leq \frac{c}{a^{\alpha }}+c(1+t^{\alpha -1})\Big( \int \lvert g_j^\sharp (t,x,v)\rvert dxdv\Big) .
\end{aligned}\]
And so,
\begin{eqnarray*}
\int |g_j^\sharp (t,x,v)|dxdv\leq \Big( \int _{\lvert v\rvert >a}f_0(x,v)dxdv+\frac{cT}{a^{\alpha }}\Big) e^{c(T+\frac{T^\alpha }{\alpha })},
\end{eqnarray*}
%\textcolor{red}{Are the previous two $\frac{c}{a^\frac{\alpha}{2}}$ correct? I get $\frac{c}{a^{\alpha}}$.}
which tends to zero when $a\rightarrow +\infty $, uniformly w.r.t. $j\geq a$. {This proves that $(f_j)_{j\in \N ^*}$ is a Cauchy sequence in $L^1([ 0,T] \times [ 0,1] \times \R ^2)$ and } ends the proof of the existence of a solution $f$ to (2.5).\\
One can similarly prove that the solution is unique and stable. The energy is non-increasing. The conservation of mass and first momentum of $f$ follow from the boundedness of the total energy. \\
\hspace*{0.1in}\\
Energy conservation will follow if the energy is non-decreasing. Taking $\psi_\epsilon=\frac{|v^2|}{1+\epsilon|v|^2}$  as approximation for $|v|^2$, it is enough to bound
\begin{eqnarray*}
\int Q(f,f)(t,x,v)\psi_\epsilon (v)dxdv = \int B\psi_{\epsilon}\Big( f^\prime f^\prime _{*}F(f)F(f_{*})
- ff_{*}F(f^\prime )F(f^\prime _{*})\Big) dxdvdv_*d\omega
\end{eqnarray*}
from below by zero in the limit $\epsilon \rightarrow 0$. Now { \cite{Lu2}}
\[ \begin{aligned}
\int Q(f,f)\psi_\epsilon dxdv
&=\frac{1}{2} \int B ff_{*}F(f^\prime )F(f^\prime _{*}\Big( \psi_\epsilon(v')+\psi_\epsilon(v'_*)-\psi_\epsilon(v)-\psi_\epsilon(v_*)\Big)dxdvdv_*d\omega \\
&\geq -\int Bff_{*}F(f^\prime )F(f^\prime _{*})\frac{\epsilon |v|^2|v_*|^2}{(1+\epsilon|v|^2)(1+\epsilon|v_*|^2)}dxdvdv_*d\omega .
\end{aligned}\]
The previous line, with the integral taken over a bounded set in $(v,v_*)$, converges to zero when $\epsilon\rightarrow 0$. In  integrating over $|v|^2+|v_*|^2\geq2\lambda^2$ , there is symmetry between the subset of the domain with $|v|^2>\lambda^2$ and the one with $|v_*|^2>\lambda^2$. We discuss the first sub-domain, for which the integral in the last line is bounded from below by
\begin{eqnarray*}
-c\int |v_*|^2f(t,x,v_*)dxdv_*\int_{|v|\geq \lambda} B \sup_{(s,x)\in [ 0,t] \times [0,1]}f^\#(s,x,v)dvd\omega\geq -c\int_{|v|\geq \lambda} \sup_{0\leq \textcolor{red}{s},x\in[0,1]}f^\#(s,x,v)dv.
\end{eqnarray*}
It follows from Lemma 4.2 that the right hand side tends to zero when $\lambda \rightarrow \infty$.
This implies that the energy is non-decreasing, and bounded from below by its initial value. That completes the proof of the theorem.     \cqfd
\\
%
%
% Proposition 4.3, i.e. asymptotic behavior of f when t tends to +infty
%
%
%An entropy connected to (1.3) is $\int s(f)dxdv$,  where
%\begin{eqnarray*}
%s(f):=f\log f +(\frac{1}{\alpha}-f)\log (1-\alpha f)^\alpha -(\frac{1}{1-\alpha}+f)\log (1+(1-\alpha)f)^{1-\alpha}.
%\end{eqnarray*}
%
\setcounter{proposition}{2}
\\
%The solution $f$ of Theorems 2.1 has its range in $]0,\frac{1}{\alpha}]$ and conserve mass and energy. This implies that $f\log f(t,.,.)\in L^1([0,1]\times \R^2)$ with $\int f(t,x,v)\log f(t,x,v)dxdv$ uniformly bounded in time.
%A simple computation shows that
%\[ \begin{aligned}
%f\log f\textcolor{red}{-f(1+ln| \frac{\alpha }{1-\alpha } | )}&\leq
%f\log f +(\frac{1}{\alpha}-f)\log (1-\alpha f)^\alpha -(\frac{1}{1-\alpha}+f)\log (1+(1-\alpha)f)^{1-\alpha}\\
%&\leq f\log f.
%\end{aligned}\]
%And so the entropy $\int s(f)(t,x,v)dxdv$ of $f$ is uniformly bounded in time. The
%entropy-entropy dissipation equation holds, {
%\[ \begin{aligned}
%\int s(f)(t,x,v)dxdv= &\int s(f_0)(x,v)dxdv-\frac{1}{4}\int _0^t\int B F(f)F(f_*)F(f')F(f'_*)(\frac{f'f'_*}{F(f')F('_*)}
%-\frac{ff_*}{F(f)F(f_*)})\\
%&(\log \frac{f'f'_*}{F(f')F('_*)}-
%log \frac{ff_*}{F(f)F(_*)})(s,x,v,v_*,\omega )dxdvdv_*d\omega ds.
%\end{aligned}\]
%Consequently, the entropy dissipation integral converges on $[0,\infty[\times [0,1]\times \R^2$. }This implies for the limit $f_\infty$ that $\frac{f_\infty}{F(f_\infty)}=e^{-\frac{(v-v_o)^2-\mu}{T}}$, which can be written
%\begin{eqnarray*}
%f_\infty=(w(e^{\frac{(v-v_o)^2-\mu}{T}})+\alpha)^{-1}\quad  \text{with}\quad  w(\zeta)^\alpha(1+w(\zeta))^{1-\alpha}=\zeta\equiv e^{\frac{(v-v_o)^2-\mu}{T}}.
%\end{eqnarray*}
%That is the equilibrium equation for anyons \cite{W}.                  \cqfd
\\
%Using Proposition 4.3, the bound $f(t,.,.)\leq\frac{1}{\alpha}-\eta_T$ for $t_m\leq t\leq T$ and $T$ given, can be improved to a global bound.\\
\\
%{\bf Corollary 4.4}\\
%{\it For the solution $f$ of Theorem 2.1, there exists $\eta_{\infty}>0$ such that $f(t,.,.)\leq \frac{1}{\alpha}-\eta_{\infty}$ for $t\geq t_m$.}\\
\\
%\underline{Proof of Corollary 4.4.}
\\
\\

\[\]


\begin{thebibliography}{99}
%
%
\bibitem
{A1} L. Arkeryd,  A quantum Boltzmann equation for Haldane statistics and hard forces; the space-homogeneous initial value problem, Comm. Math. Phys.,  298 (2010), 573-583.

\bibitem
{BBM} R. K. Bhaduri, R. S. Bhalerao, M. V. Murthy, Haldane exclusion statistics and the Boltzmann equation, J. Stat. Phys., 82 (1996), 1659-1668.

\bibitem
{BMB} R. K. Bhaduri, M. V. Murthy, M. Brack, Fermionic ground state at unitarity and Haldane exclusion statistics, J. Phys. B, 41 (2008), 115301.

\bibitem
{B} J.-M. Bony, Solutions globales born\'ees pour les mod\`eles discrets de l'\'equation de Boltzmann, en dimension 1 d'espace, in: Journ\'ees "\'Equations aux d\'eriv\'ees partielles ", Exp. XVI, \'Ecole Polytech. (1987), Palaiseau, 1-10.


%\bibitem
%{BCP} A. Biryuk, W. Craig, V. Panferov, Strong solutions of the Boltzmann equation in one spatial dimension, CRAS 342 (2006), 843-848.

%\bibitem
%{C} C. Cercignani, A remarkable estimate for the solution of the Boltzmann equation, Appl. Math. Lett. 5 (5) (1992), 59-62.
%
%\bibitem
%{errC} C. Cercignani, Errata: Weak solutions of the Boltzmann equation and energy conservation, Appl. Math. Letters 8 (1995), 95-99.
%
\bibitem
{CI} C.Cercignani, R. Illner, Global weak solutions of the Boltzmann equation in a slab with diffusive boundary conditions, Arch Rat. Mech. Anal. 134 (1996), 1-16.

\bibitem
{D} J. Dolbeault, Kinetic models and quantum effects: a modified  Boltzmann equation for Fermi-Dirac particles, Arch. Rat. Mech. Anal. 127 (1994), 101-131.

\bibitem
{EMV} M. Escobedo, S. Mischler, M. Valle, Homogeneous Boltzmann equation in quantum relativistic kinetic theory, Electronic J. Diff. Eqns., Monograph 04 (2003).

\bibitem
{H} F. D. Haldane, Fractional statistics in arbitrary dimensions: a generalization of the Pauli principle, Phys. Rev. Lett. 67 (1991), 937-940.

\bibitem
{LM} J. M. Leinaas, J. Myrheim,  On the theory of identical particles, Nuovo Cim. B N.1 (1977), 1-23.

\bibitem
{L} P. L. Lions, Compactness in Boltzmann's equation via Fourier integral operators and applications I, III, J. Math. Kyoto Univ.,34 (1994), 391-427, 539-584.

\bibitem
{Lu1} X. Lu, A modified Boltzmann equation for Bose-Einstein particles: isotropic solutions and long time behaviour, J. Stat. Phys., 98 (2000), 1335-1394.

\bibitem
{Lu2} X. Lu, On isotropic distributional solutions to the Boltzmann equation for Bose-Einstein particles, J. Stat. Phys. 116 (2004), 1597-1649.

\bibitem
{Lu3} X. Lu, The Boltzmann equation for Bose-Einstein particles: velocity concentration and convergence to equilibrium, J. Stat. Phys. 119 (2005), 1027-1067.

\bibitem
{Lu4} Lu, X., The Boltzmann equation for Bose-Einstein particles: condensation in finite time, J. Stat. Phys. 150 (2013), 1138-1176.

\bibitem
{N} L. W. Nordheim, On the kinetic methods in the new statistics and its applications in the electron theory of conductivity, Proc. Roy. Soc. London Ser. A 119, 689-698 (1928).

\bibitem
{R} G. Royat, Etude de l'\'equation d'Uehling-Uhlenbeck: existence de solutions proches de Planckiennes et \'etude num\'erique, Th\`ese, Marseille 2010.

\bibitem
{W} Y. S. Wu, Statistical distribution for generalized ideal gas of fractional-statistics particles, Phys. Rev. Lett. 73 (1994), 922-925.

%\bibitem
%{So1} Y. Sone, Kinetic Theory and Fluid Dynamics, Birkhauser, Boston 2002.

%bibitem
%*} The inequality was kindly pointed out by a referee of [1].
\end{thebibliography}
\end{document}